\documentclass{PoS}

\usepackage{url}
\usepackage{float}

\title{The Crab Nebula Spectrum at $\sim$ 100 TeV Measured with MAGIC under Very Large Zenith Angles}

\ShortTitle{Crab Nebula at $\sim$ 100 TeV by MAGIC Under Very Large Zenith Angles}
        
\author{\speaker{Michele Peresano}$^{\,,\, a, b, c}$, Razmik Mirzoyan$^{\,b}$, Ievgen Vovk$^{\,b}$, Petar Temnikov$^{\,d}$, Darko Zari\'c$^{\,e}$, Nikola Godinovi\'c$^{\,e}$, Juliane van Scherpenberg$^{\,b}$ and J\"uergen Besenrieder$^{\,b}$ for the MAGIC Collaboration \footnote{\texttt{https://magic.mpp.mpg.de/}. For collaboration list see PoS(ICRC2019)1177} \thanks{The speaker would like to thank CEA-Saclay/Irfu for its financial support. The authors on behalf of the MAGIC Collaboration: \protect\url{https://magic.mpp.mpg.de/acknowledgments_ICRC2019/}}\\
\llap{$^a$}AIM, CEA, CNRS, Universit\'e Paris-Saclay, Universit\'e Paris Diderot, Sorbonne Paris Cit\'e\\
F-91191 Gif-sur-Yvette Cedex, France\\
\llap{$^b$}Max-Planck-Institut f\"ur Physik\\
D-80805 M\"unchen, Germany\\
\llap{$^c$}Universit\'a di Udine, and INFN Trieste\\
I-33100 Udine, Italy\\
\llap{$^d$} Inst. for Nucl. Research and Nucl. Energy, Bulgarian Academy of Sciences,\\
BG-1784 Sofia, Bulgaria\\
\llap{$^e$} University of Split- FESB\\
21000 Split, Croatia\\
E-mail: \email{michele.peresano@cea.fr}}

\abstract{

The Crab Nebula was discovered as the first very-high-energy gamma-ray source by the Whipple Observatory in 1989. Thirty years after its discovery it is still the reference source and the standard candle for Imaging Atmospheric Cherenkov Telescopes (IACTs). Its spectrum has been measured from the cm radio band to energies up to tens of TeV. Some studies reported a possible but still debated cut-off in its spectrum at few tens of TeV.
The MAGIC collaboration is currently investigating the spectrum of the Crab Nebula by using the Very Large Zenith Angle observation technique. The latter provides a significantly increased collection area for energies above 10 TeV. The details of these MAGIC observations will be presented.

}

\FullConference{36th International Cosmic Ray Conference -ICRC2019-\\
		July 24th - August 1st, 2019\\
		Madison, WI, U.S.A.}

\begin{document}

\section{Introduction}
\label{sec:intro}

The Crab Nebula has a long history of observations along the entire electromagnetic spectrum \cite{crab}.
Some of the main steering arguments for its study are related to the acceleration of cosmic-rays (CRs) and their subsequent propagation in the Galaxy.
While morphological observations at lower energies indicate a complex scenario, at higher energies it seems to be possible to describe the related emission even by one-zone Synchrotron-Self-Compton models \cite{meyer}.
The extent of current VHE gamma-ray observations makes it difficult to distinguish any additional feature produced by the acceleration of hadrons \emph{in-situ}.

The extent of the CR energy spectrum makes it impossible to investigate this type of cosmic messenger with the use of just one instrument.
In almost 100 years of experiments in this direction, we have assisted to constant development in technology and measurement strategies, from balloons and satellites to ground-based facilities.
The main reason for the development of the latter - of which Imaging Atmospheric Cherenkov Telescopes (IACTs) are one of the possible implementations - has been the need to explore higher energy ranges that the one available from orbit, given the limited size of the detectors.
To further boost the energy measurement capabilities of this type of instruments, earlier works started to suggest a new technique: extending the observations at higher zenith angles. \cite{vlza1, vlza2}.
This observational method could prove to be efficient in the search for any astrophysical source capable of accelerating particles at PeV energies - i.e. PeVatrons - and shed light on the associated CR "knee" region ($\sim10^{15}$ eV).
Recent measurements by MAGIC at lower zenith angles have already brought a contribution in this direction \cite{casa}.

We report on recent developments in the observations of the Crab Nebula by the MAGIC telescopes at Very Large Zenith Angles (VLZA). This study aims to improve our knowledge of the CRs \emph{knee} by studying the VHE emission of this source towards hundreds of TeV.

\section{The MAGIC telescopes}
\label{sec:magic}

The MAGIC collaboration operates two IACTs on the island of La Palma (Canary Islands, Spain). A first telescope started to observe the VHE gamma-ray night-sky in 2004, while the second was built and commissioned in 2009, allowing stereoscopic observations. Both instruments have been upgraded in 2015 \cite{magic1}. Their hardware is now nearly identical and performances have been improved \cite{magic2}.

The MAGIC telescopes are characterized by a low energy threshold of $\sim$ 50 GeV, along with drive and alert systems build explicitly to assure competitiveness in transient studies, as recent results demonstrate \cite{txs} \cite{atel}.

Whenever possible, depending on real-time technical and observational conditions, operations of both telescopes are aided by a set of auxiliary instruments with the main purpose of monitoring the quality of the atmosphere. A micro-LIDAR system is the most important of these tools \cite{lidar}: placed on the roof of the building which hosts the on-site operations, it operates a real-time range-resolved estimation of the atmospheric transmission via backscattering of the laser light. Additional tools share this function from the center of each reflective surface of the telescopes:
\begin{itemize}
\item an optical \emph{Starguider} camera provides images of the respective Field-of-View (FoV) while serving also as pointing calibration system;
\item a CCD camera, normally used to estimate both the optical Point Spread Function (PSF) and pointing performances, but in this work also used to measure the integral transmission of the atmosphere observed through wavelength filters.
\end{itemize}

The Collaboration has recently added to this family of instruments an optical spectrograph, installed close to MAGIC 1 \cite{razmik}. Currently tested, it will measure the instantaneous transmission of the atmosphere, with higher precision than current \emph{SBig} measurements.

\section{Observing at Very Large Zenith Angles}
\label{sec:vlza}

At VLZA the propagation distance of a shower can reach $\sim<$100 km. This configuration allows to observe showers with greater Cherenkov light emission areas.
Longer propagation times along the line-of-sight (LOS) also imply that showers have to be triggered mainly by primary particles higher energies, in order to be detectable by an IACT.
The difference between the CR energy spectrum and the one from e.g. the Crab Nebula makes it clear that, as the zenith angle increases, the background rate will decrease intrinsically faster than the signal.

\begin{figure}
\begin{center}
\includegraphics[width=.95\textwidth]{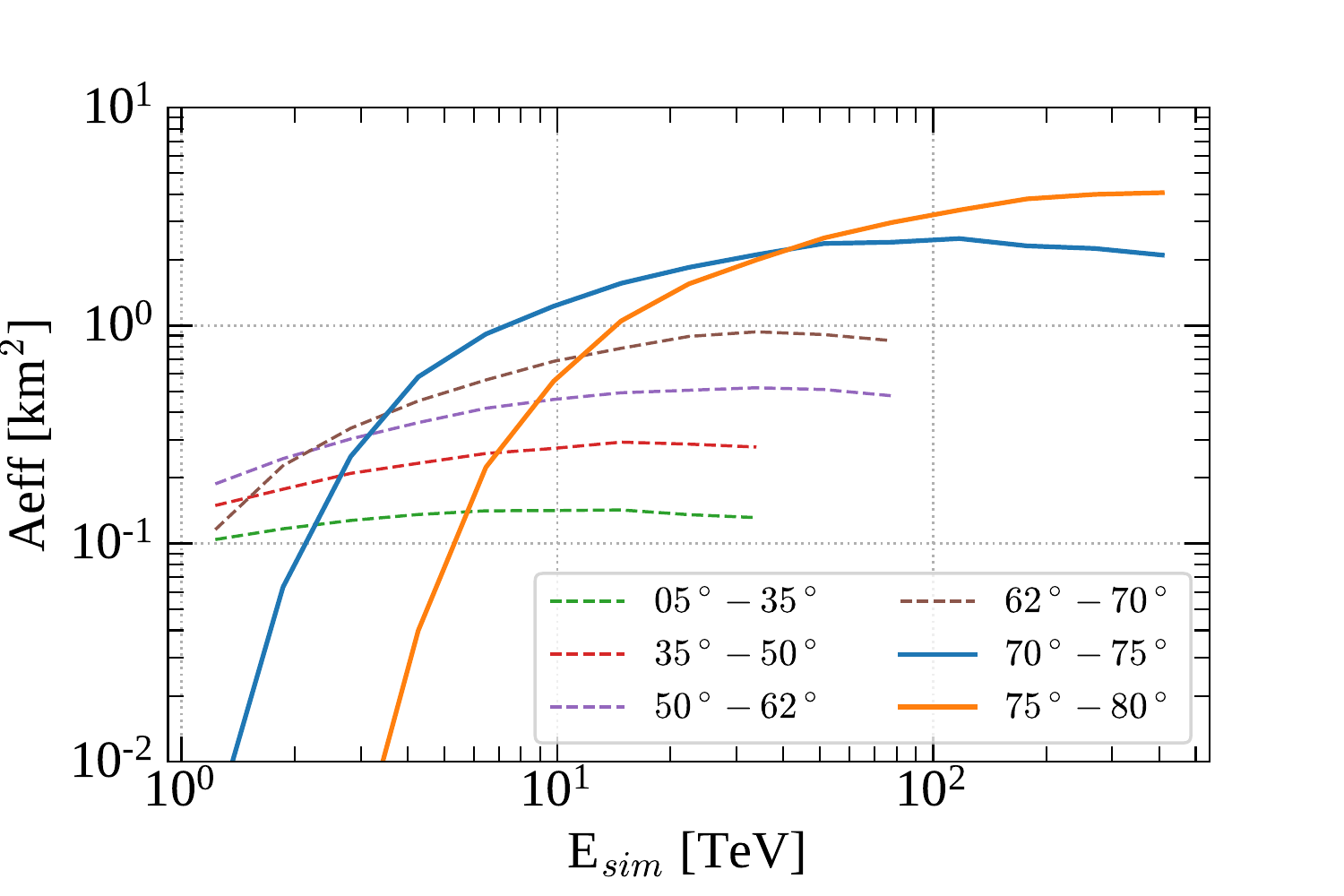}
\caption{Collection area of the MAGIC telescopes as a function of simulated energy; the color code describes different zenith configurations. Data from $70^{\circ}$ onwards have been simulated specifically for this work.}
\label{fig:coll}
\end{center}
\end{figure}

The principal gain of IACT observations at greater zenith angles is thus a better collection area at the price of a higher energy threshold, as shown in Fig.\ref{fig:coll}.
For a more detailed description of how the MAGIC telescopes applied this observational technique to unprecedentedly large zenith angles, please see also \cite{vlzatech}.

\section{Observations and data analysis at VLZA}

The Crab Nebula is still observed to emit a stable $\gamma$-ray flux even at tens of TeV \cite{magic2}: this could make it a first choice for calibration of IACT instruments even at VLZA.

Up to now, $\sim 50$ hours of good quality data have been accumulated by observing the Crab Nebula between $70^{\circ}$ and $80^{\circ}$ with an energy threshold in the range $1 \sim$ 10 TeV.

\subsection{Data quality selection}

The usual data quality process characterizing MAGIC data at lower zenith angles could not be applied with the same level of precision.
The process developed for this study makes use of cross-correlations between low-level quantities, both from some of the MAGIC subsystems (camera and drive) and from a few auxiliary instruments (LIDAR and \emph{Starguider} cameras).

The overall sample of data is initially reduced using two global cuts:
\begin{itemize}
\item $\gtrsim70\%$ of atmospheric transmission from LIDAR data within a distance of 12 km,
\item a cut on the Direct Current (DC) measured by each MAGIC camera photo-multipliers.
\end{itemize}
The former ensures that the showers possess a minimum quality and comply with usual lower-zenith observations.
The latter selects only data compatible with a \emph{dark} night-sky-background (NSB), i.e. DC $< 2.2 \mu A$ as measured from the MAGIC-1 camera \cite{moon}.

\begin{figure}
\begin{center}
\includegraphics[width=.95\textwidth]{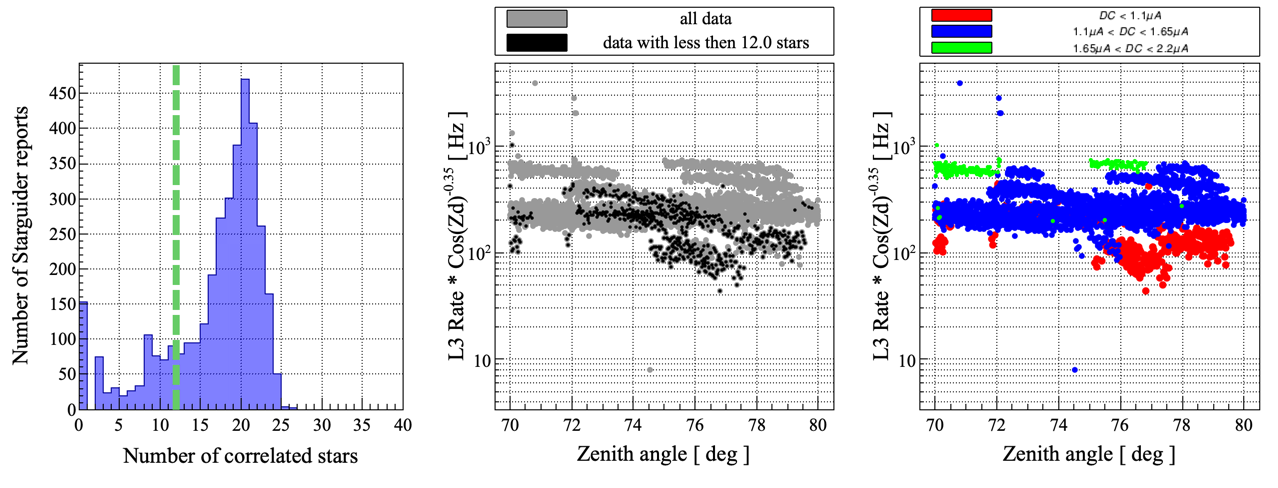}
\caption{Example of the quality of a subset data from this work. \emph{Left panel}: application of the cut on number of stars in the FoV; \emph{Middle panel}: effect of this cut on the stereo trigger rate as a function of zenith angle; \emph{Right panel}: stereo trigger rate as a function of zenith angle divided in narrow DC levels.}
\label{fig:vlza_data_quality}
\end{center}
\end{figure}

With reference to Fig.\ref{fig:vlza_data_quality}, a minimum amount of stars are required to be recognized in the wobble-dependent \cite{wobble} FoV by each \emph{Starguider} camera (left panel). The effect of this cut is then studied on the event stereo trigger rate as a function of zenith angles (middle panel).
If some absorbing material is present between one telescope and a shower along the LoS, then the local number of stars is expected to decrease correspondingly, resulting in a correlated decrease of the rates (black points in middle panel).
The opposite effect can be produced when other light sources become relevant (e.g. the Moon). In this case the absorber could scatter also this light in all directions: trigger rates would increase while the number of stars in the FoV would be unchanged.
The DC levels have to be monitored simultaneously (right panel) because an evolving NSB could easily mimic this effect. With a clear sky the presence of the Moon will still increase the number of stereo-triggered events (light green data in right panel).

\subsection{Atmospheric monitoring for energy estimation via optical spectrography}

To correct for bias effects in the energy estimation due to bigger amounts atmosphere, it has been necessary to improve the study of the observing conditions along with ongoing specific treatments of the systematics. This involves auxiliary instruments and limited resolution of the Monte-Carlo simulations.
For zenith angles $\lesssim 70^{\circ}$ the main instrument for this purpose would have been the micro-LIDAR, but for VLZA observations the use of this auxiliary tool is not sufficient anymore. Given its limited emission power, the estimation of the absorption of Cherenkov light cannot be probed along the entire shower propagation distance at VLZA. This in turn limits the reliability of the energy estimation.

\begin{figure}[H]
\begin{center}
\includegraphics[width=.6\textwidth]{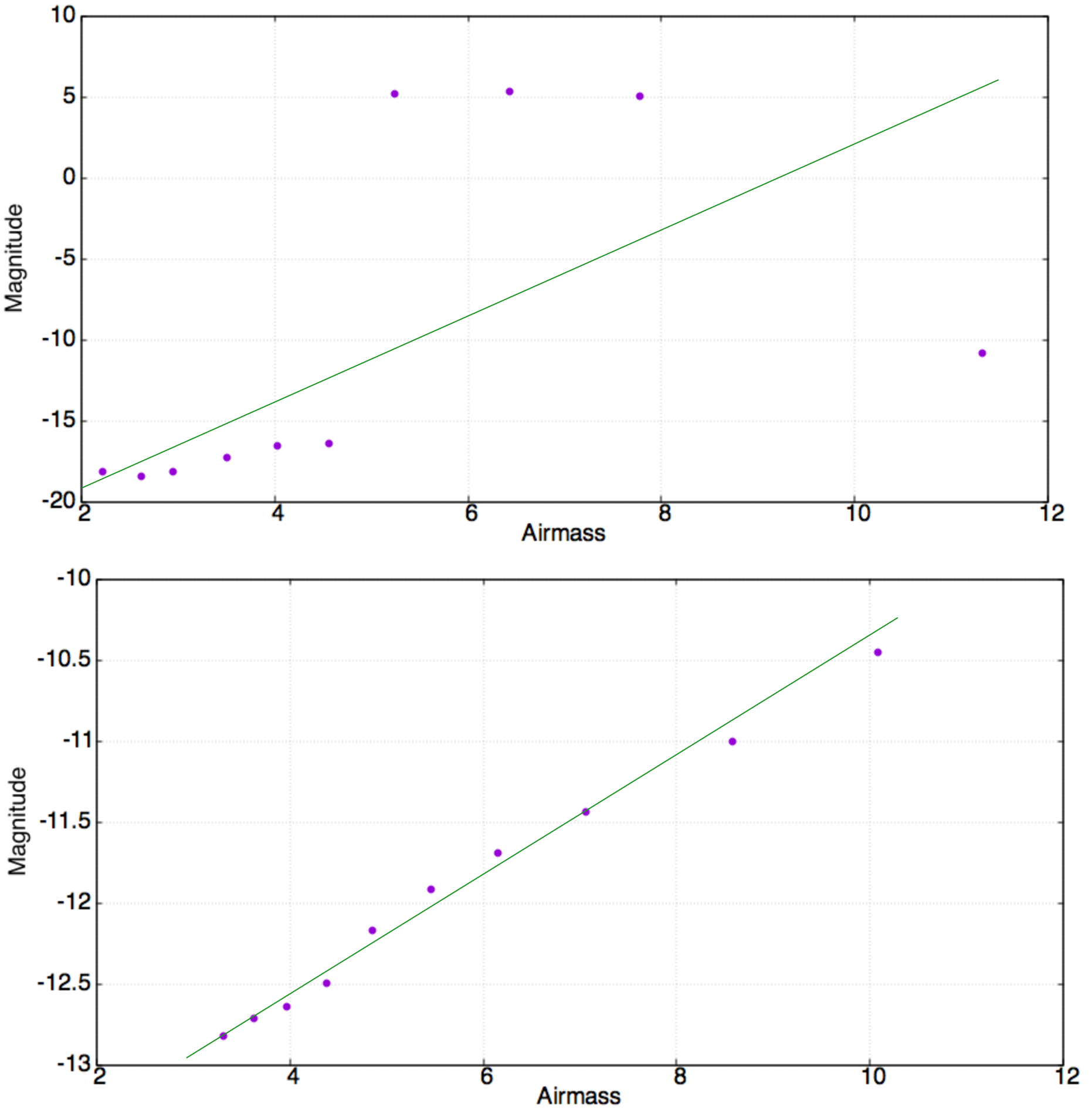}
\caption{Example showing the difference at VLZA between a clear night (lower panel) and one in which a certain amount of absorbing material is present along the LoS (upper panel). The slope of the fitted linear relation (green) represents the correction factor to be applied on the shower-dependent data, while its intercept is an estimate of the known star's magnitude at the top of the atmosphere.}
\label{fig:sbig}
\end{center}
\end{figure}

For this reason, we applied and developed an alternative technique for calibration and correction of atmospheric data at VLZA.
This work makes use of spectrographic images obtained with different optical filters with at a minute-scale period.
By measuring the variation of magnitude from bright stars used as a reference it is possible to estimate the integral atmospheric absorption suffered from VLZA showers along the LoS \cite{razmik} \cite{me} (see Fig.\ref{fig:sbig} for an example).

The calibration of these measurements is done by measurements the same patches of sky during clear nights, with auxiliary information at lower zenith angles from the LIDAR and instruments external to MAGIC which are useful for multiple observational facilities in La Palma.

\section{Conclusion}

The MAGIC telescopes are applying the Very Large Zenith Angle observation technique at unprecedentedly large zenith angles to observe the Crab Nebula at the highest energies.
Dedicated MonteCarlo data have been produced in order to simulate the instrument responses from $70^{\circ}$ to $80^{\circ}$.
Reliable results were obtained thanks to an improvement of the usual procedures regarding atmospheric monitoring and data quality.
A detailed study of the relevant systematics is ongoing. Even if the main MAGIC hardware performance will likely not change at VLZA, most of the current contribution to this type of uncertainty is expected to come from the limited knowledge of how the showers develop and are imaged at such large zenith angles.

By increasing the collection area of the telescopes, the VLZA technique will likely extend the current capabilities of the MAGIC telescopes to hundreds of TeV requiring less observational time than previous facilities of the same kind. 
These efforts will allow the MAGIC telescopes to investigate the presence of PeVatron candidates before the complete development of the future Cherenkov Telescope Array (CTA), in which this technique could be further boosted.

\end{document}